# Preprint Serious Game Based Dysphonic Rehabilitation Tool

Zhihan Lv
FIVAN, Valencia, Spain
Email: lvzhihan@gmail.com

Chantal Esteve
FIVAN, Valencia, Spain
Email: chantal@fivan.org

Javier Chirivella
FIVAN, Valencia, Spain

Pablo Gagliardo
FIVAN, Valencia, Spain
Email: pablog@fivan.org

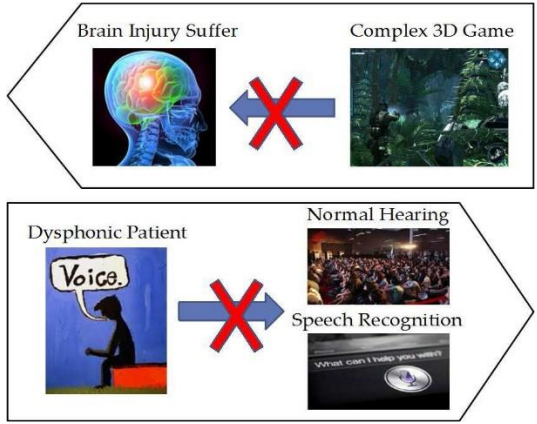

Fig. 1. The limitation of the comprehension and expression of dysphonic.

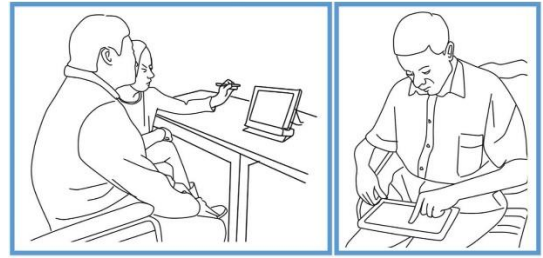

Fig. 2. Scenarios. Left: At clinic; Right: At home.

*Abstract*—The purpose of this work is designing and implementing a rehabilitation software for dysphonic patients. Constant training is a key factor for this type of therapy. The patient can play the game as well as conduct the voice training simultaneously guided by therapists at clinic or exercise independently at home. The voice information can be recorded and extracted for evaluating the long-time rehabilitation progress.

## I. INTRODUCTION

Contemporary healthcare systems are continuously under pressure to reduce costs. Using video game as the vision feedback has been already demonstrated as a useful Human-Computer-Interaction (HCI) approach for rehabilitation assistive method [6]. Some previous research have proved the usability of the game in speech therapy [16], for children with autism [4], and parkinsons disease patients [5]. In our research we employ video game to evaluate the efficacy of the patient voice exercise, which can be considered as the long-time continued evaluation of the severity of dysphonia. We define our research as a biometric information management system, in which the biometric refer to voice pitch specifically, while serious games are employed as the user interface.

In clinical, dysphonia is measured using a variety of tools that allow the clinician to see the pattern of vibration of the vocal folds, principally laryngeal videostroboscopy. Besides invasive or drug based treatments, effective logopedic treatments have been proved [14]. The dysphonic voice can be hoarse or excessively breathy, harsh or rough, but some kind of phonation is still possible, therefore, the clear speech recognition is not expected. The patients usually have inabilities of pronunciation of the high pitches. The hopeful rehabilitation tool is able to retrieve the quantified long-time voice rehabilitation information. The voice exercise can not only improve the pronunciation but also enhance the swallowing muscle (Velum and Pharynx) of the patient.

This research is from practical need. In one side, most of our patients are acquired brain injury suffer. The complex 3D game cannot be understood by them, even most of them haven't played with 3D immersive games using new interaction devices. So the rehabilitation tool should have friendly user interactive experience and be able to be flexible customized by therapists according to the specific condition of patients. In another side, the patient voice cannot be understood by untrained people with normal hearing or standard speech recognition technology. So the specific pitch estimation algorithm and reasonable game logic should be designed and developed.

## II. SYSTEM

According to the therapists suggestions and our clinical observation, we designed an serious game based rehabilitation tool with a user interface. The software is able to estimate the basic information of patient voice (i.e. pitch (Mel)). We have got the voice sample during the preliminary clinical test. The voice sample is used for evaluating seven pitch estimation algorithms, they are also known as pitch detection or pitch recognition algorithms. The most suitable pitch estimation algorithm for Dysphonic patient voice will be selected by the evaluation. Further more, we plan to measure and compare the patient's emotion after the traditional treatment to that after serious game based treatment.

Our rich clinical experience gives rise to unique game design principle. The entertainment is significant for the designed game, because it can stimulate the initiative and enhance the motivation of the patient so that the patients are willing to pay more effort to the rehabilitation exercise. 2D game is enough attractive, because the patients we have met often confuse the 3D manipulation rule. Besides, some of the patients have disability of 3D perception. Therefore, aiming to our practical demand, 2D UI is prefered and easier to navigate, traverse and engage, which satisfies criteria that arise from the nature of patient's spatial perception [22]. As suggested by the therapist, pitch (Mel) is the main controlable factor, thus we uses it as the manipulating tool which function is equal to the joystick in the classic video game. Some other factors are also considered, such as Phonation time (ms), pitch change (Mel), Duration (s), Reaction time (ms). Built based on the considered factors, the configuration includes some essential parameters which can modify the difficult level of the games. The parameters contain sensitivity, x spread, y spread, incoming speed, voice maintenance, session duration. The parameters are available to modify by the level editor. The designed functions can provide immediate evaluation feedback to therapists and patients.

A new version software adapted to the tablet with normal micphone is on-going developed. This version is getting rid of kinect and doesn't need the supervision of therapist. It could be used for rehabilitation exercise at home. The patient can choose the suitable version as willing. For both version, the biometric data are collected after the game and uploaded to the server side for long-time rehabilitation data analysis, and further generating the rehabilitation suggestions. The implied gamification feature during this process is that the patient try the best to follow the game rule.

III. CONCLUSION

Virtual reality has been widely applied in many fields range from everyday entertainment and communication [8] [23] to traditional research fields such as geography [15] [12] [17], climate [18] and biology [19] [3], clinical assist [9]. 'Intuitive', 'Entertainment' and 'Incentive' are three main factors of our design principle during our gradual and iterative research process. Some novel interaction approaches are considered to be integrated in our future work [7] [11] [10]. The new network data management algorithm [20] [21] , smart grid system [2] [1] [24] [13], will be also considered.

ACKNOWLEDGMENT

The authors would like to thank Sonia Blasco, Vicente Penades and Oktawia Karkoszka. The work is supported by LanPercept, a Marie Curie Initial Training Network funded through the 7th EU Framework Programme (316748).